\documentclass[12pt,a4paper,dvips,epsfig]{article} 
\usepackage{graphicx}
\usepackage{subfigure,rotating}

\begin{document}
\begin{titlepage}
\begin{flushright}\vbox{\begin{tabular}{c}
IMSc-2000/09/51 \\
September 2000\\
hep-ph/0009216\\
\end{tabular}}\end{flushright}

\begin{center}
{\bf \large Dileptons from $\eta_c$ in Nucleus-Nucleus collisions} 
\end{center}
\bigskip
\begin{center}
Ramesh Anishetty and Rahul Basu \footnote{e-mail :ramesha,rahul@imsc.ernet.in}
\\  
{\em The Institute of Mathematical Sciences, \\ 
Chennai(Madras) - 600 113, INDIA.} \\ 
\end{center}
\bigskip
\begin{abstract}
Preliminary estimates suggest that excess dimuon production with
invariant mass in the range 1.5 -- 2.5 GeV in nucleus-nucleus collisions 
can be explained on the basis of  
$\eta_c$ production. This appears to be consistent with all the peripheral and
central collision data with various nuclei such as S-U at 200 GeV/nucleon 
except for the central collision data on Pb-Pb at 158 GeV/nucleon. 
Some explanations based on glueball production for Pb-Pb data are discussed.
\end{abstract}
\end{titlepage}
Over the past decade many different experiments in Nucleus-Nucleus
(A-A) collisions \cite{hist} have consistently shown excess dilepton production
with dimuon
invariant mass in the range 1.5-2.5 GeV for $\mu^+\mu^-$  
pairs.
These data sets were interpreted by appropriately scaling proton-nucleus 
(p-A) data. Essentially in all these data sets the dilepton sources are either
Drell-Yan pairs or decays of $J/ \psi$ or $D{\bar D}$ . The bulk of the
data appears to agree in general by consideration of these sources alone.
However there is a significant departure between the observed dilepton
pairs and the theoretical estimate based on the above sources, in the 
intermediate mass
range (IMR) (i.e. $\mu^+\mu^-$ invariant mass in the range 1.5-2.5 GeV).

In the literature there have been  several attempts to explain this
discrepancy. Some of these explanations are based on decrease in $\rho$ meson 
mass due to thermal effects in $e^+e^-$ data \cite{lkb}, D-rescattering 
\cite{lw}, 
enhanced $D{\bar D}$ production, in-flight $\pi^+\pi^-$
decaying to $e^+e^-$ \cite{tserruya},
and so on. Many of these
physical processes suggested as an explanation are interesting in their own 
right but in all of these the  
explanation for excess dileptons in the IMR is at best partial 
and generally tend to be relevant in a regime different from the IMR. Fireball
hydrodynamics with adjustable parameters however seem to contribute in the IMR
regime in Pb-Pb at 158 GeV/nucleon central collisions \cite{rs}. 

Present data for di-leptons from p-A and A-A collisions over the entire
kinematic regime upto 5 GeV agrees with the conventional QCD explanation
in terms of Drell-Yan process and vector mesons {\em except for the IMR region 
mentioned above}.  The overall
picture involving charm quarks \cite{kharzeev} is that when 
there is sufficient energy 
exchange in a
collision, protons have non-negligible charm content ($c{\bar c}$
pairs), and substantial high
energy gluons which in turn can decay to $c{\bar c}$ pairs. These 
$c{\bar c}$ pairs
occassionally form bound states such as $J/\psi$ by emitting a soft
gluon to maintain color balance, or can further polarize $u{\bar u}$ or 
$d{\bar d}$
from the surrounding medium to form $D{\bar D}$ pairs. Although the present
theoretical understanding cannot predict absolute numbers for these
processes it is possible to check the consistency of this picture with 
various p-A
and A-A data. By and large the data agrees with various quantitative
checks. 

This picture however, also suggests that other charm meson bound states such
as $\eta_c,\psi^\prime$ and $\chi^\prime$'s are produced as well. The relative 
abundance 
of each of these mesons is expected
to be constrained by their sizes. In particular the larger the size of
the charm meson the less likely it is to  get formed in the hadronic plasma
\cite{kharzeev}.
This expectation is clearly borne out by the data, {\em viz.} in any
experimental setup $J/\psi$ which is a 1S orbital state is produced about
100 times more than $\psi^\prime$  which is a 2S orbital state. Both these
resonance peaks are clearly visible in the dimuon data.

In this paper we are concerned about the production of the $\eta_c$
meson. This is a 1S orbital state and is expected
to have the same size as a $J/\psi$; furthermore it has almost the same
mass. They differ only in their spin and hence it is expected that in
any collision where $c{\bar c}$ quarks are produced these can form $\eta_c$ 
with about $1/3$ probability as $J/\psi$. In fact, any suppression
mechanisms \cite{ms,ks} due to the hot hadronic medium will equally affect 
both the
mesons. Consequently it is fair to expect that in any of the experimental
setups the production cross sections of $J/\psi$ and $\eta_c$ are similar.
The $\eta_c$'s once produced will typically decay into lighter hadrons, 
which makes the 
direct detection of $\eta_c$ virtually impossible. However there is a
small (estimable) cross section for it to decay into $\gamma\mu^+\mu^-$
or $\gamma e^+e^-$. The relative decay probabilities of $\eta_c$  to 
$\gamma\mu^+\mu^-$ and that of $J/\psi$ to $\mu^+\mu^-$ is essentially 
determined by
the electromagnetic interaction of the charm quark. 
\begin{figure}
\begin{center}
 \includegraphics[width=4.8in]{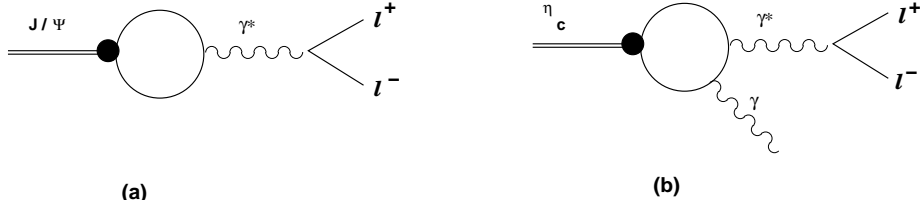}
\end{center}
\caption[]{Production of dileptons from a)$J/\psi$ and b)$\eta_c$, where
the blob represents the bound state wave function.}
\label{fig:prod}
\end{figure}
The reason for this is clear from Fig. 1 which shows 
$J/\psi$ decay in
Fig 1a and that of $\eta_c$ in Fig 1b where $\gamma^*$ in turn
decays into a $\mu^+\mu^-$ pair.
If we
restrict ourselves to large invariant mass for 
the $\gamma^*$ ( $>$ 1 GeV in Fig1b), it is reasonable to expect the 
charm quark
inside the loop to be almost free. In other words, if the $J/\psi$ spin 
dependent wave function
is $\gamma_\mu \phi$ and the $\eta_c$ wave function is $\gamma_5 \phi$ where 
$\phi$ refers to the remaining spin and spatial part of the wave function,
then (ref.  Fig 1a. and Fig 1b.) when the mass of $\gamma^*$ is above 1 GeV,
we can take  $\phi$ to be well appproximated by free
quark and anti-quark propagators upto an overall constant. Hence in the
ratio of these processes this constant is irrelevant. In Fig 1b. we
integrate the kinematic space for physical $\gamma$ and find the $\eta_c$
contribution to $\mu^+\mu^-$ pair is given by \cite{iz} (for $M^2\leq
M_\eta^2$)
\begin{equation}
\frac{dN}{dM}=\frac{1}{3}N_{\psi}\frac{1}{60}\frac{\alpha}{\pi}\frac{m_c^2}
{m_{\psi}^2}\Big[\frac{M_{\eta}^2}{M^2}-1\Big]
\frac{\vert\int_0^1 {dx\over x}\ln\frac{m_c^2-x(1-x)M_{\eta}^2}
{m_c^2-x(1-x)M^2}\vert^2}{\vert\int_0^1 dx x(1-x)\ln\frac{m_c^2-x(1-x)
M_{\psi}^2} {\Lambda^2}\vert^2}
\end{equation}
where $m_c$ is the mass of the charm quark and $M_\eta$ and $M_\psi$ are the 
masses of the $\eta_c$
and $ J/\psi$ mesons. $\Lambda$ is a cut-off parameter which regulates the
logarithmic divergence in Fig.1a. Here $N_\psi$ is the number of $J/\psi$
events per unit mass around the $J/\psi$ peak. In reality, in the p-A
and A-A experiments, the $J/\psi$ peak is broadened due to detector
resolution. Consequently, we interpret $N_\psi$ as the total number of
$J/\psi$ events under the broadened peak. 

Now we would like to remark on the limitations of this preliminary analysis.
In any experimental set up the detection of $\mu^+\mu^-$ is limited by
various cuts - in particular, rapidity and Collins-Soper \cite{abreu}
angle cuts. It is with
these constraints that the experiment determines $N_\psi$, the number of $J/
\psi$ events seen. $J/\psi$ undergoes a two body decay to dimuons and hence
will have different acceptance ratios to that of the three body decay of 
$\eta_c$.
These differences can be taken into account in detail, by doing a Monte Carlo 
simulation
of the experiment using PYTHIA, for instance. In this preliminary analysis we
shall ignore these details.
Consequently we are
assuming that the acceptance ratios of $J/\psi$ and $\eta_c$ events are about
the same. It is known, for example, that the experimental acceptance
ratio is about 15\% for $J/\psi$ events. For $\eta_c$ we expect that
when most of the energy is carried away by the dimuons alone and in
addition, we also do not detect the real photon, it is possible that 
the acceptance ratio
for $\eta_c$ may not be very different. However, for sufficiently small
invariant mass ($<$ 1 GeV) this assumption will fail. 
In spite of ignoring these
details, we feel that the qualitative features of our analysis can be of
importance in understanding the excess dimuon production in the IMR.

The data on S-U collisions at 200 GeV/nucleon \cite{abreu} is one of the best 
studied
experimentally in terms of statistics and comparisons between central vs 
peripheral collisions.
Furthermore, most of the features of the earlier experiments with different
nuclei are essentially contained,
with better statistics, in this data. We therefore take the S-U data as
the typical example for doing our analysis and
take $\Lambda$ = .85 GeV to reproduce the excess dimuon events in the
central S-U collisions as shown in Fig.2a. Using the same value for $\Lambda$
we then check with the peripheral collision data, as shown in Fig.2b.
Both appear to be reasonably satisfactory. It shows, for example, that in the
IMR nearly 10 - 14 \% 
of events (depending on peripheral or central), appear to come from 
$\eta_c$.
Taking $m_c=1.6$ GeV, it turns out that the logarithm in the denominator in
Eq.(1) 
is very sensitive to the choice
of $\Lambda$. This is only to be expected since the $J/\psi$ mass is very 
close to $c{\bar c}$ threshold. We further notice that after choosing 
$\Lambda$, the
exact choice of $m_c$ in the range 1.1 to 1.8 GeV is insignificant to our
estimates. 
\begin{figure}[htb]
\begin{center}
 \includegraphics[width=2.6in]{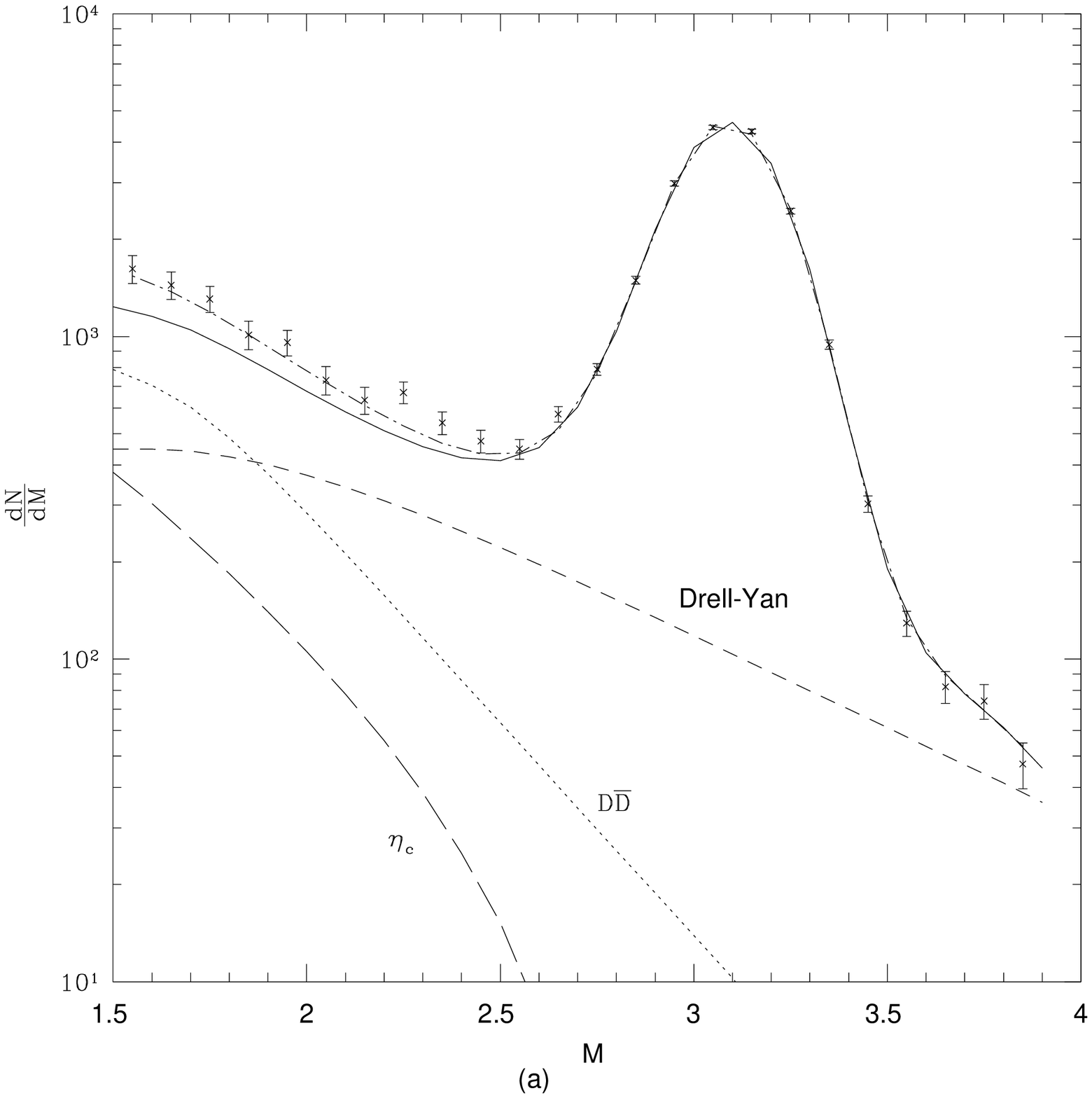}\hfill 
\includegraphics [width=2.6in]{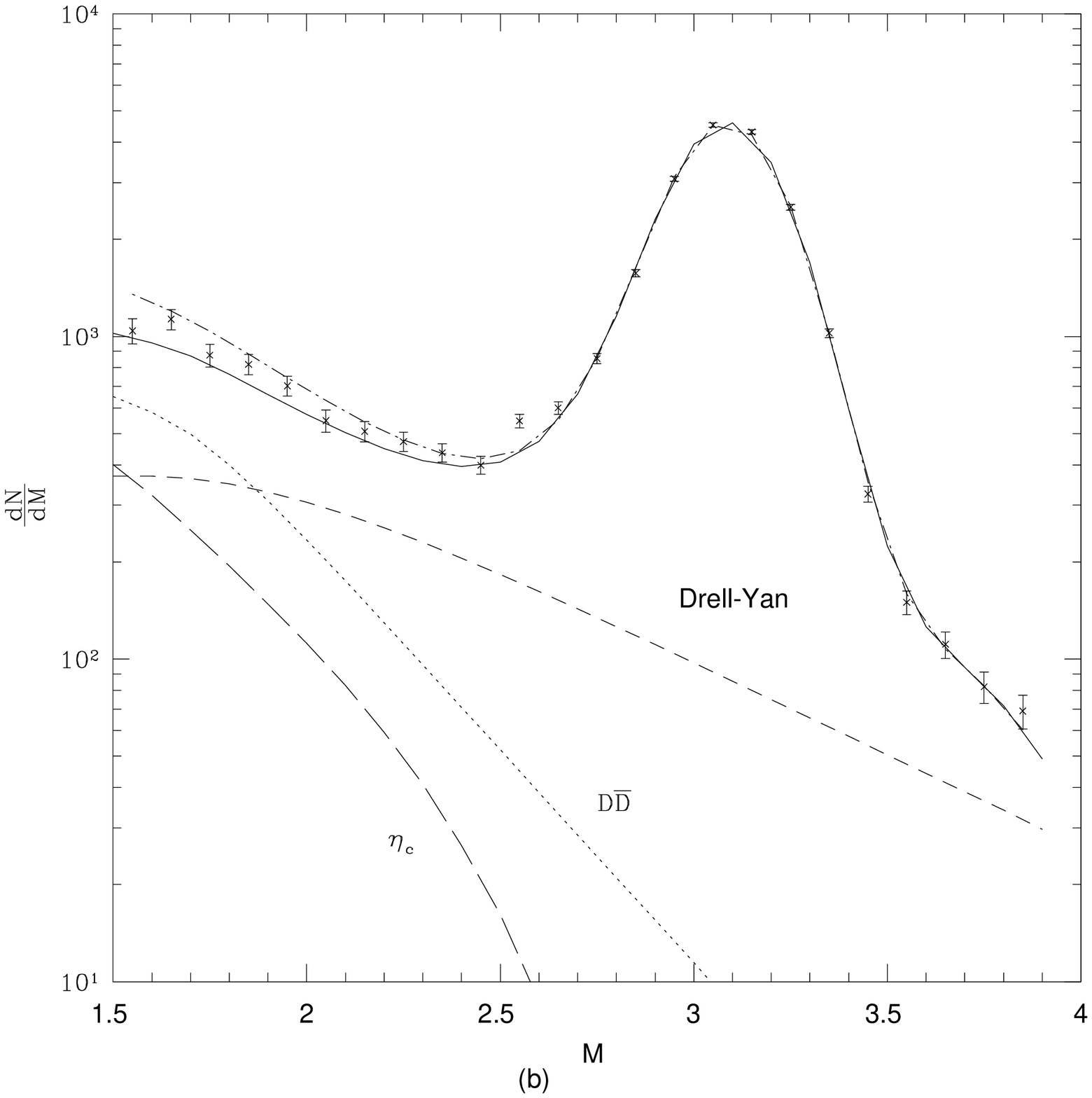}
\end{center}
\caption[]{S-U data - a)central and b) peripheral collisions. The solid
line is the total contribution excluding $\eta_c$\protect\cite{abreu}. The 
dot-dashed line is the total contribution including $\eta_c$. The individual
contributions are also shown.}
\label{fig:S-U}
\end{figure}

For the chosen values of $\Lambda$ and $m_c$, we can check the p-A data against 
the corresponding $N_\psi$. We find that in these data sets the $\eta_c$ 
contribution, although substantial in IMR, is still less than the background 
as estimated by the experimental groups \cite{abreu}. In fact, both in
p-A and A-A the $\eta_c$ contribution is less than the $D{\bar D}$
contribution (see Fig. 2). From \cite{abreu} it is clear that in the p-A
data the $D{\bar D}$ contribution is an order of magnitude lower than
the combinatorial background. Hence the $\eta_c$ contribution is also
much below the background. 

Next, data for Pb-Pb at 158 GeV/nucleon data is fitted 
with the same parameters $m_c$ and $\Lambda$ obtained earlier from the
S-U data. $N_\psi$ here is taken from the Pb-Pb data (for central or
peripheral collisions as the case may be). This is shown in Fig. 3a and
3b (central and peripheral). 
Peripheral data is reasonably well accounted for with our $\eta_c $ production
mechanism. (This peripheral data can be sensitive to the experimental cuts 
and therefore this agreement should be re-examined in a more detailed
analysis incorporating all the experimental cuts). 
However in the
central collision data Fig.3a the discrepancy between our explanation and the
data is still about 40\% in the IMR. This discrepancy in the total number of
events is rather less sensitive to the experimental cuts. Consequently it is
fair to infer that our explanation in terms of $\eta_c$ production for central 
collisions of Pb-Pb at 158 GeV/nucleon is not complete.  
\begin{figure}[htb]
\begin{center}
 \includegraphics[width=2.6in]{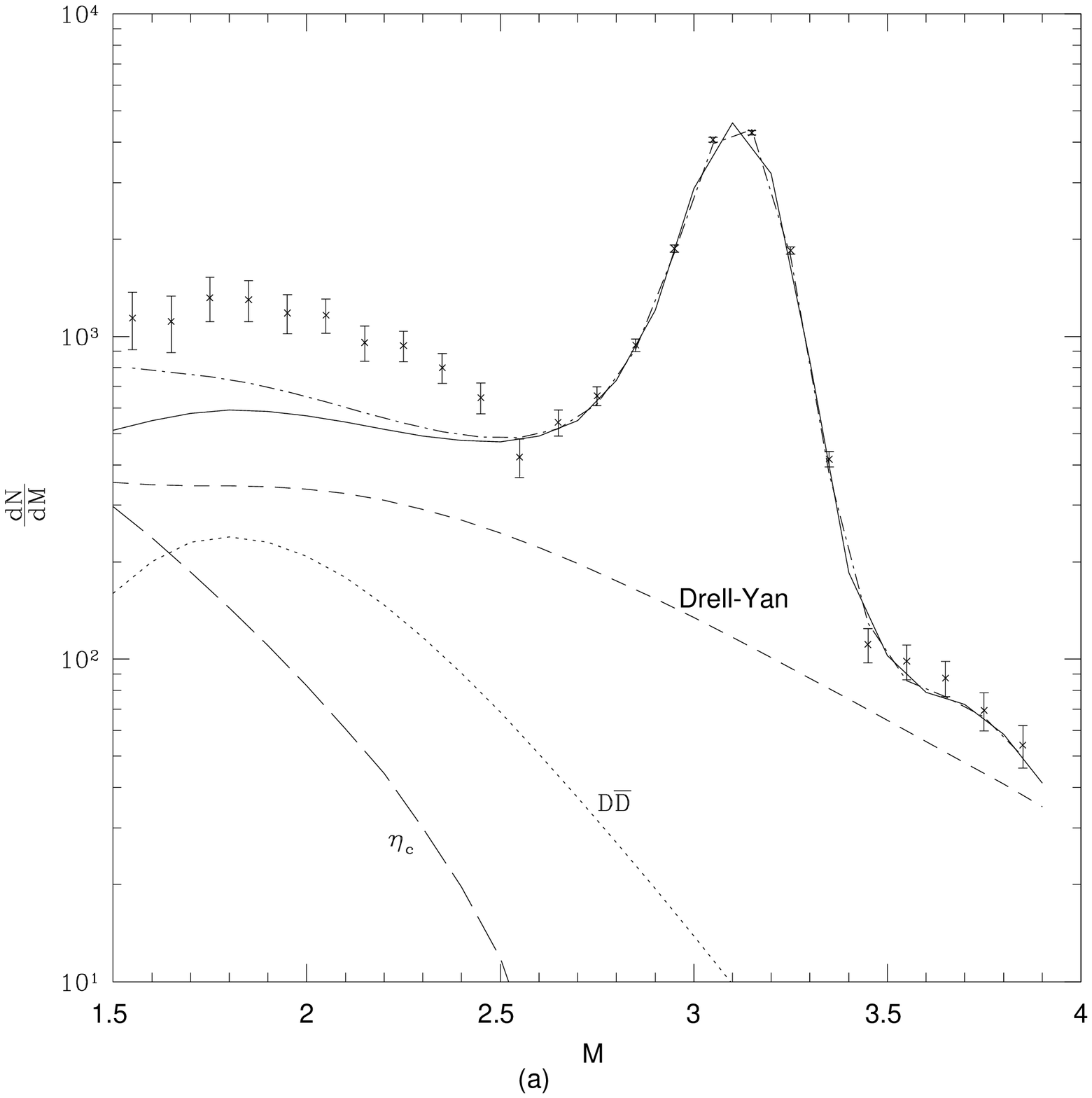}\hfill
\includegraphics [width=2.6in]{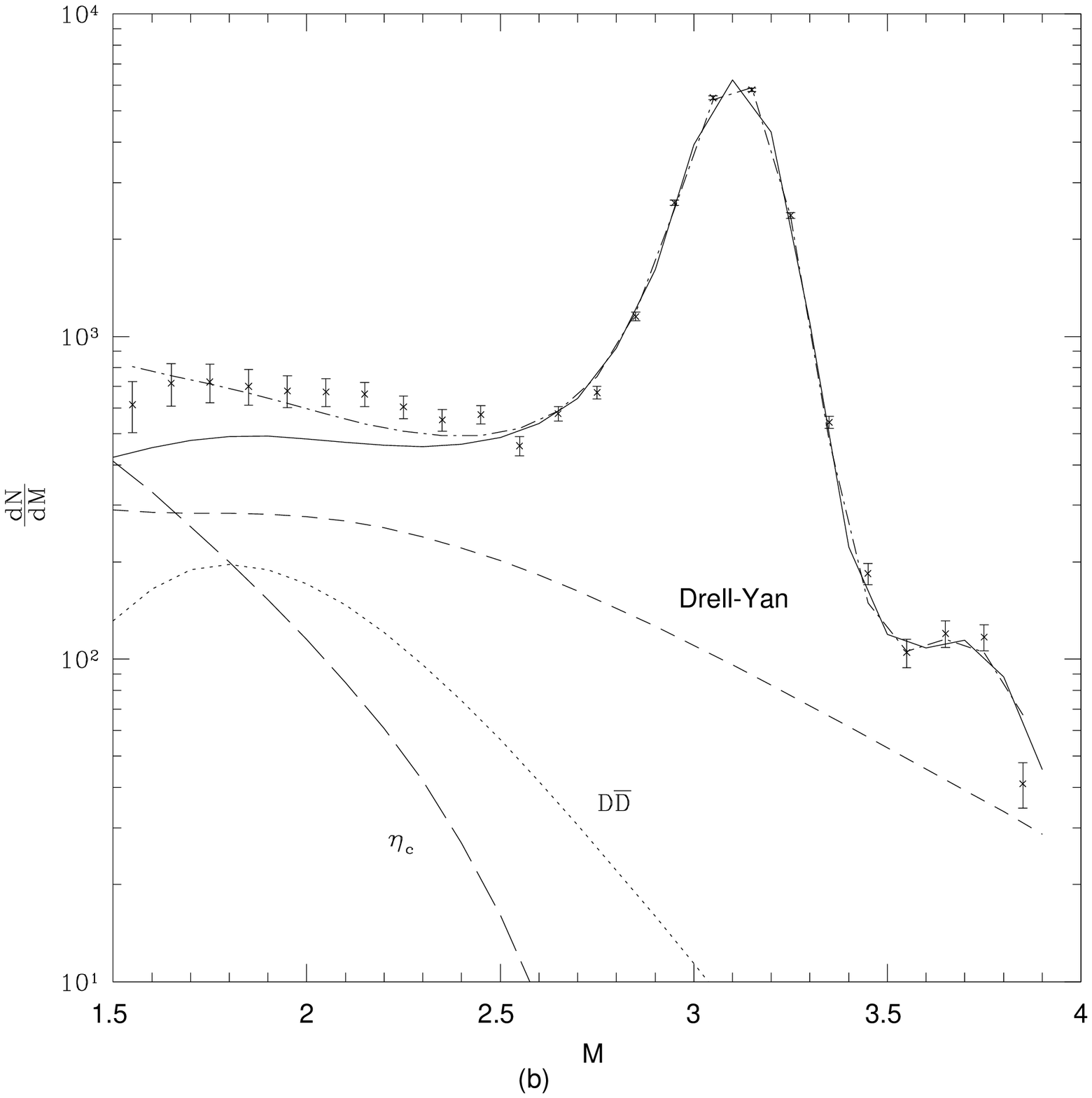}
\end{center}
\caption[]{Pb-Pb data - a)central and b) peripheral collisions.
The solid line is the total contribution excluding $\eta_c$. The 
dot-dashed line is the total contribution including $\eta_c$. The individual
contributions are also shown.}
\label{fig:Pb-Pb}
\end{figure}

For $e^+ e^-$  data we have not redone this analysis since the
corresponding $N_\psi$ is not quoted by the experimental group. It is clear
from the analysis however, that there can be substantial contribution here as 
well from $\eta_c$.

The net result of this analysis is that the $\eta_c$ contribution to IMR of 
dimuon events is substantial and can fit the data satisfactorily in all A-A
data excepting central Pb-Pb at 158 GeV/nucleon. There is substantial other evidence
which suggests that in this data the hadronic plasma may have undergone QGP 
transition; consequently the hydrodynamic evolution parameters can play an
important role \cite{rs}. 

Finally we would like to speculate over what other candidates from QCD 
phenomenology
may play a role in the dilepton data. It has long been suspected that
QCD has glueballs, although experimentally there is as yet no strong evidence 
- only some candidate events. In the context of A-A collisions, we can expect 
that a gluon-rich medium
either hadronic or QGP in nature can produce such particles. These in turn can 
form glueballs of various spin, which again decay into
standard mesons and baryons. This makes the detection of glueballs extremely 
tricky.
Occassionally due to electromagnetic interaction through quark loops they do
decay into photons and lepton pairs. Estimates of these processes is
essentiallly hampered by our lack of understanding of glueball production.
Noting that most of A-A collision data other than Pb-Pb are well explained
by our $\eta_c$ scenario, it would not be out of place to surmise  
that perhaps some glueballs may be produced after QGP in
Pb-Pb collision. Let us now expand on this scenario.

{}From Fig.3a, after accounting for $\eta_c$, the major discrepancy in the data 
occurs in the narrow region 1.7 - 2.3 GeV. Scalar or pseudoscalar glueballs,
lighter than 2 GeV cannot contribute to the above region as their decay
goes through a $\gamma^*\gamma$ process. Thus only 
heavier than 3 GeV glueballs will contribute over the IMR.
Since our present theoretical prejudices coming from either lattice 
\cite{patel,bs,svw} or sum rule
techniques \cite{svz, novikov} suggest that the lightest glueballs are 
around 1.5 - 2 GeV, these
light scalar or pseudoscalars cannot contribute to the discrepancy in Fig.3a.
On the other hand the range 1.7- 2.3 GeV is reasonably narrow, which
suggests that perhaps this
is due to the $1^-$ vector glueball with a mass of about 2 GeV decaying to
dimuons. The production rate of this vector glueball is not estimable. If
we presume that all the discrepancy is to be accomodated by the $1^-$ vector
glueball then we can infer from Fig. 3a that about $3-4\times 10^3$ dimuon 
events are due to this glueball. 

Finally a word of caution. It is clear that in central Pb-Pb at
158 GeV/nucleon some interesting different physics such as the 
hydrodynamic fireball expansion model as envisaged by \cite{rs} may be at work. 
In such a case our 
association of all the extra events to glueballs alone would be incorrect.
It would be worthwhile therefore if there were other corroborative evidence 
in the data to support glueball production.  

Keeping this in mind, we can speculate that in RHIC, where the QGP phase
is produced and lasts longer, the effects of glueballs would be
enhanced. The net effect of this in the dimuon spectrum in the IMR
would be an enhancement of the bump in that region, thereby increasing
the discrepancy between the data and present explanations by much more
than 40\%. 

To summarize, the standard QCD explanation for the dimuon spectrum
naturally involves the $\eta_c$ contribution as well. Interestingly,
this contribution can be self-consistently estimated and the inclusion
of this effect explains the various experimental data. However, central
Pb-Pb data which is believed to undergo perhaps a QGP phase transition
continues to show a discrepancy which perhaps needs explaining through
a qualitatively new mechanism.

\bigskip
\noindent{\bf Acknowledgements}: We are grateful to E. Scomparin for sending 
us the data sets for S-U and Pb-Pb scattering and for useful comments.
We would like to thank the anonymous referee for clarifying certain
matters and thereby helping us present the paper with more clarity.


\bigskip

\begin{thebibliography}{99}
\bibitem{hist}
I. Ravinovich for CERES Coll., Nucl. Phys. {\bf A638}, 159C (1998);
G. Agakichiev for CERES Coll., Nucl. Phys. {\bf B422}, 405 (1998);
M. Masera for HELIOS3 Coll., Nucl. Phys. {\bf A590}, 93C (1995);
A. De Falco for NA38 Coll., Nucl. Phys. {\bf A638}, 487C (1998);
E. Scomparin for NA50 Coll., J. Phys. {\bf G25}, 235 (1999);
A. Drees, Nucl. Phys. {\bf A610}, 536C (1996).
\bibitem{lw}
Z. Lin and X. N. Wang, Phys. Lett. {\bf B444}, 245 (1998).
\bibitem{tserruya} 
I. Tserruya, LANL Archives nucl-ex/9912003 and references therein.
\bibitem{lkb}
G. Q. Li, C. M. Ko and G. E. Brown, Phys. Rev. Lett. {\bf 75}, 4007
(1995); Nucl. Phys. {\bf A606}, 568 (1996).
\bibitem{rs}
R. Rapp and E. Shuryak, Phys. Lett. {\bf B473}, 13 (2000).
\bibitem{kharzeev}
D. Kharzeev, Nucl. Physics {\bf A638}, 279C (1998) and references
therein.
\bibitem{ms}
T. Matsui and H. Satz, Phys. Lett {\bf 178}, 416 (1986).
\bibitem{ks}
D. Kharzeev and H. Satz, Phys. Lett. {\bf B366}, 316 (1996)
\bibitem{iz}
C. Itzykson and J.-B. Zuber, Quantum Field Theory, McGraw Hill (1980)
page 552.
\bibitem{abreu}
M. C. Abreu {\it et al.}, E. Phys. Jour. {\bf C14}, 443 (2000).
\bibitem{patel}
A. D. Patel {\it et al.}, Phys. Rev. Lett. {\bf 57}, 1288 (1986)
\bibitem{bs}
T. H. Burnett and S. R. Sharpe, Ann. Rev. Nucl. \& Part. Sci. {\bf 40}.
327 (1990) and references therein.
\bibitem{svw}
J. Sexton, A. Vaccarino and D. Weingarten, Nucl. Phys. Proc. Suppl. {\bf
B42}, 279 (1995).
\bibitem{svz}
M. A. Shifman, A. I. Vainshtein and V. I. Zakharov, Nucl. Phys. {\bf
B147}, 385 (1979), {\it ibid.} 448 (1979).
\bibitem{novikov}
V. A. Novikov {\it et al.}, Nucl. Phys. {\bf B191}, 301 (1981).
\end{thebibliography}
\end{document}